\documentclass[10pt, conference]{IEEEtran}
\usepackage{hyperref}
\usepackage{graphicx, subfig}
\usepackage{amsmath, amssymb}
\usepackage{algorithm}
\usepackage{algpseudocode}
\usepackage{amsmath}
\graphicspath{ {images/} }
\usepackage[T1]{fontenc}
\usepackage[utf8]{inputenc}
\usepackage{caption}
\usepackage{todonotes}
\usepackage{comment}

\makeatletter
\begin{document}
\title{A Software-Defined QoS Provisioning Framework for HPC Applications}
%Authors

\author{\IEEEauthorblockN{Neda Tavakoli}
\IEEEauthorblockA{Dept. of Computer Science\\
		Texas Tech University\\
		Lubbock, TX, USA 79409\\
		Email: neda.tavakoli@ttu.edu}
\and
\IEEEauthorblockN{Yong Chen}
\IEEEauthorblockA{Dept. of Computer Science\\
		Texas Tech University\\
		Lubbock, TX, USA 79409\\
				Email: yong.chen@ttu.edu}}
\maketitle

\begin{abstract}
With the emergence of large-scale data-intensive high-performance applications, new I/O challenges appear in the efficient management of petabytes of information in High-Performance Computing (HPC) environments. Data management environments must meet the performance needs of such applications, represented by various Quality-of-Service (QoS) metrics such as desired bandwidth, response time guarantee, and resource utilization. Traditional high-performance management platforms are facing considerable challenges regarding flexibility, as well as the need to address a variety of QoS metrics and constraints. To tackle these challenges, a Software-Defined approach is considered promising, and various prototypes have already been deployed in Cloud-based data centers.
In this paper, we investigate the idea of utilizing a software-defined approach to provide I/O QoS provisioning for HPC applications. We identify the key challenges towards the high degree of concurrency and variation in HPC platforms, and propose a series of novel designs into the general software-defined approach in order to deliver our goal. Specifically, we introduced a borrowing-based strategy and a new M-LWDF algorithm based on traditional token-bucket algorithms to assure a fair and efficient utilization of resources for HPC applications. Due to the lack of software-defined frameworks in current HPC platform, we evaluated our framework through simulation. The experimental results show that our strategies make a significant improvement upon the general HPC frameworks and lead to clear performance gain for HPC applications.
\end{abstract}
 
\section{Introduction}
\label{sec:2.Introduction}
%outline: Motivation: problem, existing solutions, problem of existing solution
Application I/O performance is a critical design concern for HPC frameworks. However, the advent of petascale computing for modern HPC applications is leading to enormous concurrencies in HPC frameworks. This new trend of growth in size, complexity of data and concurrencies for HPC applications has created tremendous challenges for the application I/O performance. Because of this, new HPC management frameworks are in demand for addressing concerns originating from the performance needs of such applications. 
Traditional HPC management frameworks face considerable challenges regarding the performance requirements of such applications (e.g., response time, resource utilization, desired bandwidth, and other QoS metrics). However, satisfying the performance needs of applications (i.e. QoS requirements) is critical to support efficient and scalable performance. To the best of our knowledge, no flexible and programmable HPC framework exists that ensures application QoS requirements. 

%Outline: Problem statement
Motivated by these considerations, we design and produce an HPC framework that supports QoS provisioning for HPC applications using a software-defined approach. We call this framework the \emph{Software-Defined QoS Provisioning} (SDQPro) framework for modern HPC Applications. Our goal in this effort is to provide an efficient, programmable, and flexible software-defined HPC framework that satisfies the performance needs of modern HPC applications. In fact, this paper answers the question of which applications should wait and which ones can use the shared HPC framework to meet all applications QoS requirements. The goal of QoS in this context is to allocate bandwidth among multiple HPC applications in such a way as to provision the desired bandwidth for each of them. In our framework, if any changes occur in the applications QoS requirements, we do not necessarily need to re-configure thousands of devices; instead, by re-configuring a centralized software, other devices will be automatically notified. 
%
 %Outline: Prior work, why prior work is inadequate.
The literature offers numerous solutions that address applications I/O requirements (e.g., see~\cite{ chang2008bigtable, krauter2002taxonomy, chervenak2000data,shoshani1998storage}). For example, many efforts towards these goals have been done in~\cite{ chang2008bigtable}, by designing the \emph{Bigtable}, which provides the dynamic control over data management. However, this dynamic control is not as flexible as our software-defined approach. 

%Key ideas of the paper
In this paper, we identify and formally define the problem of QoS provisioning for HPC applications using a software-defined approach. To this end, two major software-defined components, the \emph{data plane} and the \emph{control plane} ~\cite{lantz2010network,mckeown2008openflow,kim2013improving} are added to the traditional HPC storage management system to adaptively deal with I/O operations of today's applications. Typically, the data plane is used to collect I/O requests from entire applications and to classify them into appropriate dedicated queues based on their I/O headers. The centralized control plane monitors and controls the entire framework to make decisions of whether the applications are able to obtain their specified bandwidth; or still get appropriate shares of the bandwidths even the storage systems are overloaded. A token-bucket and borrowing-based algorithm is used to assure a certain level of resources per application. Such a framework enables the HPC management system to address the high-bandwidth, dynamic nature of today's applications and involves managing storage devices by setting policies of how and in which order they can be used to ensure the desirable performance needs per application. Intuitively, the centralized control plane, based on the global view of the entire system, makes the decisions of which applications should wait and which one can use the shared storage systems to meet application QoS requirements.

Finally, we evaluate our framework within a simulated environment and use synthetic benchmarks to validate our proposed framework. The evaluation results demonstrate that by using the software-defined framework, applications are either able to obtain their specified bandwidth if possible, or still get appropriate shares of the bandwidths, even the HPC storage systems are overloaded. 

The contributions of this research study are threefold: 
\begin{itemize}
	\item Our primary contribution is to identify the architecture of an HPC framework with the ability of QoS provisioning for HPC applications using a software-defined approach;
	\item  The second contribution of this work is achieved from the performance gains offered by the centralized control plane~\cite{kim2013improving} which motivated us to use this approach for storage management frameworks;
	\item Our third contribution is to develop the policy enforcement component to dynamically configure the overall framework to the QoS requests of applications.
\end{itemize}

The rest of this paper is organized as follows: Section~\ref{sec:3.Background and Challenges} presents motivation and background needed for this study, typical HPC architectures and their existing challenges, and description of software-defined QoS-aware frameworks, along with more detailed discussion of their major components. In Section~\ref{sec:7.Design}, our Software-defined QoS- aware algorithm for HPC applications is discussed in detail. Section \ref{sec:8.Evaluation}  describes the evaluation and results of this study.  In Section~\ref{sec:4.RelatedWorks}, we discuss the related works, and then we conclude our approach in Section~\ref{sec:9.Conclusion}.

\section{Background and Terminologies}
\label{sec:3.Background and Challenges}
This section introduces required background and terminologies for this study. At first, a typical HPC architecture is described as a fundamental HPC framework used for this study. Then, existing challenges in dealing with application I/O requests in typical HPC frameworks are explained in detail. Afterwards, our software-defined storage framework, its terminologies, and major components are explained in detail. 
 
A typical HPC framework has three major components: compute nodes, a communication network, and distributed storage servers. The compute nodes are typically responsible for running and computing multiple processes from different parallel applications.  Another HPC component is the communication network, which is used to connect compute nodes to the distributed storage servers. Distributed storage servers store file data on one or more object storage target (OST) devices. We use this architecture as a fundamental framework for our design.   

\subsection{Applications I/O in HPC}
As each application requires I/O requests from different shared storage servers, the major challenge in the typical HPC architecture and implementation is to deal with applications' I/O performance. Because of this, application I/O plays a critical role on the performance of current generation HPC systems. In addition, the performance mismatch between the compute nodes and applications I/O requests of today's HPC systems has made I/O a critical bottleneck for HPC applications. This paper attempts to bridge this increasing performance gap by adding dedicated software to adaptively deal with applications I/O. 

\subsubsection{Challenges}
The typical HPC framework has two major challenges while handling application I/O requirements, and this study attempts to address both of them. These challenges are listed below.

\begin{figure}[htbp]
	\begin{center}
		{\includegraphics[width=2.4in]{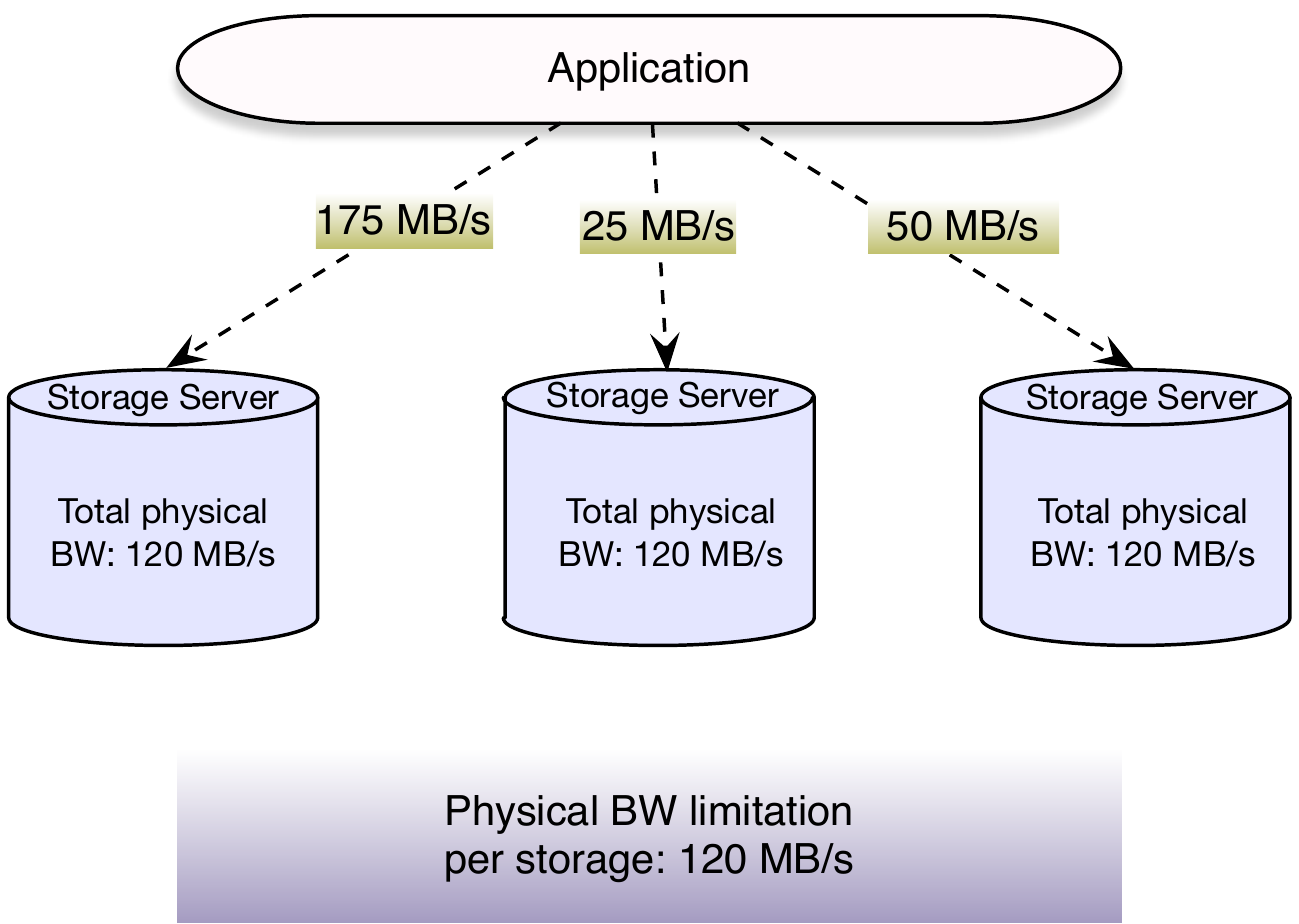}}
		\caption{Unbalanced I/O requests}
		\label{fig2}
	\end{center}
	\vspace{-1em}
\end{figure}

\paragraph{Unbalanced I/O requests}
HPC applications often perform I/O in an unbalanced way, which can cause a problem for the fair sharing of bandwidth among shared storage servers. For example, Fig.~\ref{fig2} illustrates an example of this situation. For simplicity, only one application is considered in this example. It issues unbalanced I/O requests from different compute nodes hitting different storage servers. As the figure shows, this application issues a total amount of 300 MB/s of unbalanced I/O requests from the storage servers while the total physical bandwidth (BW) limit for each storage is 120 MB/s. 

The ideal case is that the I/O requests evenly distribute on storage servers such that each of them has 100 MB/S of requests. But in this example, I/O requests are unevenly issued on storage servers. Based on these physical limitations, the HPC framework could only serve requests at 275 MB/s, even though other servers actually have unused capacity for this application. This issue is similar to the problem of bursty traffic in networking areas, which can be alleviated using token bucket algorithms~\cite{tang1999network, shenker1997general}. To address such an issue, we propose a token bucket borrowing model algorithm described in the Section~\ref{sec:7.Design}.

\paragraph{Physical limitations}
Each storage server has a physical bandwidth limitation. This limitation can be reached due to unbalanced I/O requests from a single application or due to many concurrent data accesses from multiple applications. To address this issue, a borrowing model algorithm is used.  

The rest of this section introduces the software-defined storage framework and its various components.

\subsection{Software-Defined Storage Framework in HPC systems}
\label{sec:6.SDS}
Software-Defined Networking (SDN)~\cite{lantz2010network,mckeown2008openflow,kim2013improving} is a new networking paradigm that decouples the logic required to control a piece of hardware from the underlying hardware itself. Therefore, the logic is then controlled by software. SDN is compromised of two major components, the data plane and the control plane. The control plane can be programmed via an interface (e.g., OpenFlow~\cite{shalimov2013advanced}, etc) which is used to monitor and control the hardware (i.e., data plane). This decoupling promises to dramatically simplify the management of the entire network. 

\begin{figure}[htbp]
	\begin{center}
		{\includegraphics[width=3.6in]{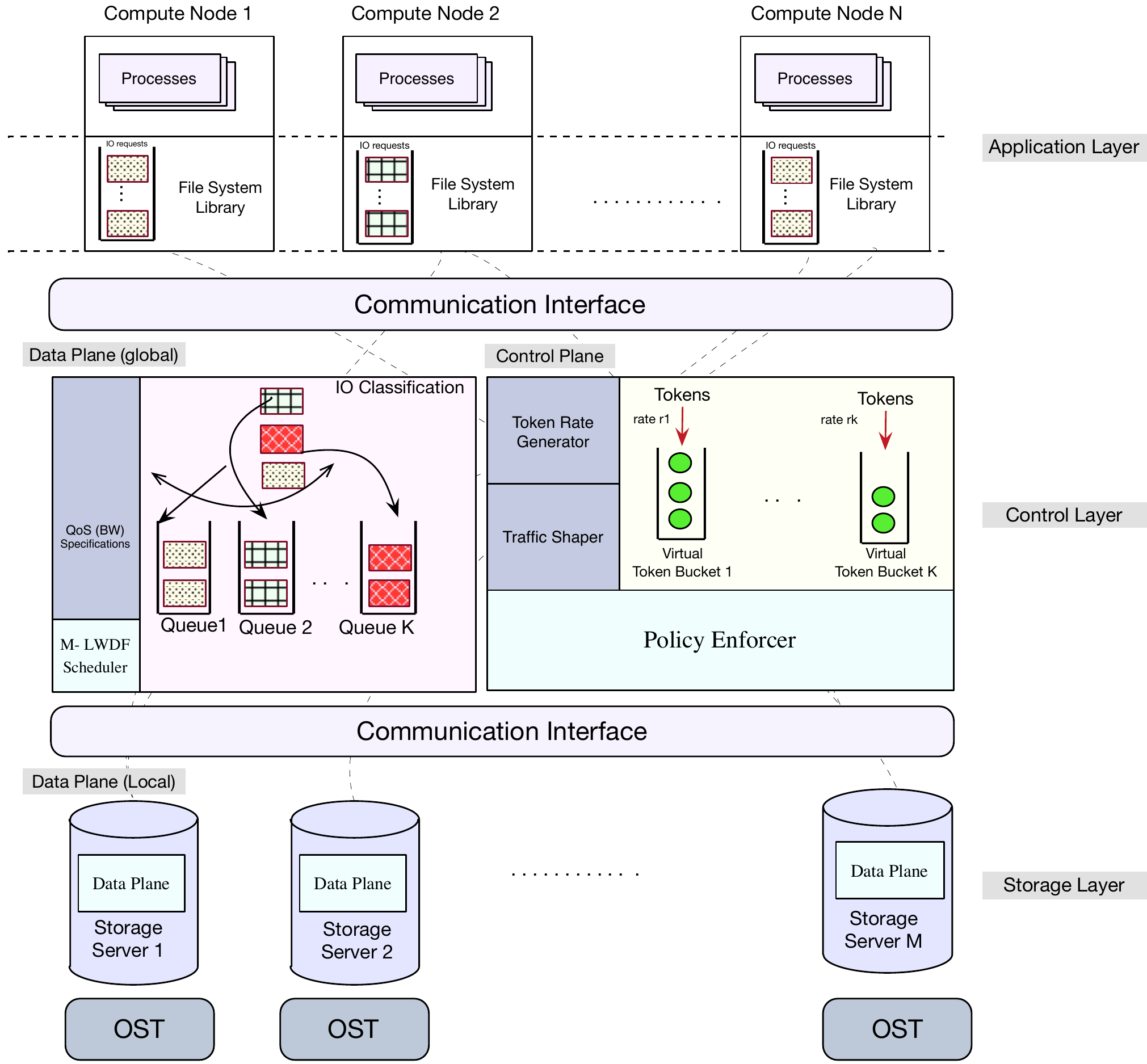}}
		%\caption{QoS-Aware architecture using software-defined approach.}
		\caption{A software-defined QoS provisioning architecture for HPC applications}
		\label{fig1}
	\end{center}
\end{figure}

Fig.~\ref{fig1} illustrates our proposed software-defined storage framework. It is comprised of three major layers; \emph{application layer}, \emph{control layer}, and \emph{storage layer}. The first layer, the application layer, is formed of compute nodes running parallel applications such that I/O requests per application are collected in queues inside the compute nodes. The second layer, the control layer (which is the main contribution of this paper), is used to monitor and control the entire framework. The control layer introduces a new layer in the I/O software stack which interposes a piece of software above the file system layer but below the rest of the I/O software stack. Finally, the storage layer is used to manage the entire storage system. The storage layer directly communicates to the control layer using a communication interface to program the data plane located inside the storage devices. The rest of this section explains each layer in detail.
\subsection{Application layer}
This layer is compromised of compute nodes running HPC applications in a parallel fashion. The application may have many processes, which are distributed among compute nodes and most of their processing time is devoted to I/O operations. As shown in Fig.~\ref{fig1}, the I/O requests (i.e., read or write) are stored on dedicated queues assigned to each of the compute nodes (i.e., one queue is assigned per compute node). These I/O requests then must be moved to the storage layer to be executed. To this end, a communication interface between the application layer and the control layer used to send the dedicated compute node queues to the control layer. Then, these node-level queues are classified to separate queues per application. Finally, the communication layer between the control layer and the storage layer is used to move the I/O requests to proper storage devices. 

\subsection{Control layer}
This layer consists of three major components; \emph{distributed data planes}, the \emph{centralized control plane}, and the \emph{communication interfaces} defined between the layers.  

\paragraph{Distributed data planes}
As shown in Fig.~\ref{fig1}, distributed data planes are compromised of two-level data planes. The first layer, called the \emph{global data plane}, consists of only one data plane. The second layer data plane, called the \emph{local data plane}, has a pool of local data planes located on the storage layer and distributed between entire storage servers. In general, data planes, whether they are global or local, have the same architecture comprising of three components; \emph{I/O classification}, \emph{QoS specification}, and \emph{M-LWDF scheduler}. Each component is defined as follows: First, the I/O classification component consists of a queue per application which classifies the applications' I/O requests based on their I/O header. Because of this classification, I/O requests of the same application are put to the dedicated queue for that application. Another component is \emph{QoS (BW) specification} which collects desired QoS (bandwidth) requirements per application. The final component is the M-LWDF scheduler, which was explained in Section~\ref{sec:3.Background and Challenges}, which satisfies each individual QoS (bandwidth) requirement on a single shared storage server. In the following, the roles of two level data planes are explained in detail. Then different components of the data plane are discussed. 
\begin{enumerate}
	\item The global data plane collects I/O requests from all applications. These requests are then classified to dedicated queues assigned per application. Classification is done based on the I/O headers such that requests with the same header belong to the same application and will be put to the same queue.
	\item As shown in Fig.~\ref{fig1}, the pool of local data planes is located on the storage layer (i.e., one local data plane for one storage server). Local data planes have the same functionality as the global one. However, the main difference between them is that the global data plane deals with I/O requests of all applications. In contrast, each local data plane only handles I/O requests issued to it.
\end{enumerate}

\paragraph{Centralized control plane}
\label{controlP}
This is a logically centralized component, called a controller, which dictates the overall storage behaviour. It is the "brain" of the framework where control logic is implemented. This paradigm brings several benefits when compared to traditional methods. First, it is much easier to introduce new storage behaviour in the storage servers through a software program. Second, it introduces the benefits of a centralized approach to storage configuration, as opposed to distributed management: operators do not have to configure all storage servers individually to make changes in storage behaviour, but they may instead make decisions in a logically singular location, the controller, with global knowledge of the storage system state.
Th control plane consists of four components which are described as follows.
First, the \textit{Token Rate Generator} component which communicates with the QoS (BW) Specification component in each data plane to sync the requested bandwidth specification of each application. Based on this information, it generates a token rate per application. Another control plane component is the \emph{Traffic Shaper} which equally distributes generated \emph{tokens} into the queue of each corresponding application in the data plane. 
A token is a conceptual data structure representing the permission of performing IO requests. Only if a queue is holding enough tokens, will it be served by a storage server. The other control plane component, the \textit{Virtual Token Buckets}, learns the token rate from the token rate generator and generates corresponding tokens per application. A bucket (buffer) is used per application to keep track of its corresponding tokens (note that the number of generated tokens per application is different based on their I/O needs). Last, the \textit{Policy Enforcer} is used to deliver policies which meet the QoS requirements. For example, a  policy can be \texttt{<app-1, rate=100 MB/s>} which means application1 can be forced to have a bandwidth of 100 MB/s, no matter how much bandwidth was actually requested. The policy enforcer provides a huge degree of flexibility when configuring the entire system. It will apply the configuration policies to potentially hundreds of storage servers.

\paragraph{Communication interface}
As shown in Fig.~\ref{fig1}, two communication interfaces are used for this study. One is located between the application layer and the control layer, and the other one is placed between the control layer and the storage layer. The communication interfaces enable a direct interaction among the layers. For instance, the communication interface located between the application layer and the control layer permits the control layer to collect all application I/O requirements. The other communication interface enables a direct interaction between the control plane and the data planes. Because of this, the data planes can be programmed by the control plane to meet QoS requirements.

\subsection{Storage layer}
The storage layer is comprised of storage servers with a data plane on each storage unit, and \emph{OST (object server target)}. The local data plane runs on each storage server for I/O classification and bandwidth shaping for each application. As shown in Fig.~\ref{fig1}, a communication interface is used to allow the storage layer to directly interact with the control layer, and vice versa.

\section{Software-defined QoS-aware algorithm for HPC applications}
\label{sec:7.Design}
The architecture shown in Fig.~\ref{fig1} is used for QoS provisioning. This section introduces the algorithm used for this purpose in detail.  Motivated by the work done in~\cite{andrews2001providing}, we propose a new software-defined QoS-aware algorithm to ensure application bandwidth requirements are met in HPC systems. As shown in~\cite{andrews2001providing},  the M-LWDF scheduling is used in conjunction with the token bucket algorithm, to satisfy a certain amount of bandwidth for applications that are using a single shared resource.

The rest of this section is organized as follows: first we explain the token bucket algorithm used in storage context. Then, we discuss the proposed policy enforcer and the extended M-LWDF algorithm. After that, we discuss the token-bucket algorithm in conjunction with the Extended M-LWDF used for this study. Finally, we explain how our proposed solution works in the general case.

\subsection{Token bucket algorithms in storage servers}
Token bucket algorithms ~\cite{tang1999network, shenker1997general} are widely used in ATM networks to manage and shape the burstiness of network traffic. Essentially,  the token bucket allows bursts below a regulated maximum rate. Tokens are generated at a fixed rate over time (i.e., one token every $\Delta t$), and each token represents the permission to send a specific amount of data.  In this paper, this  method is used to handle the unbalanced I/O requests of HPC applications. We use the token bucket algorithm to limit the storage bandwidth usage per application. As shown in Fig.~\ref{fig1}, the token bucket algorithm is implemented in the control plane. It assigns tokens per application in order to limit the storage BW usage for the application. The process of generating and distributing tokens was explained in detail in Section~\ref{sec:6.SDS}. However, as shown in~\cite{andrews2001providing}, the token bucket algorithm alone is not enough to ensure the applications QoS requirements are met, rather the M-LWDF must be used in conjunction with it. 

\begin{figure}[htbp]
	\begin{center}
		{\includegraphics[width=2.3in]{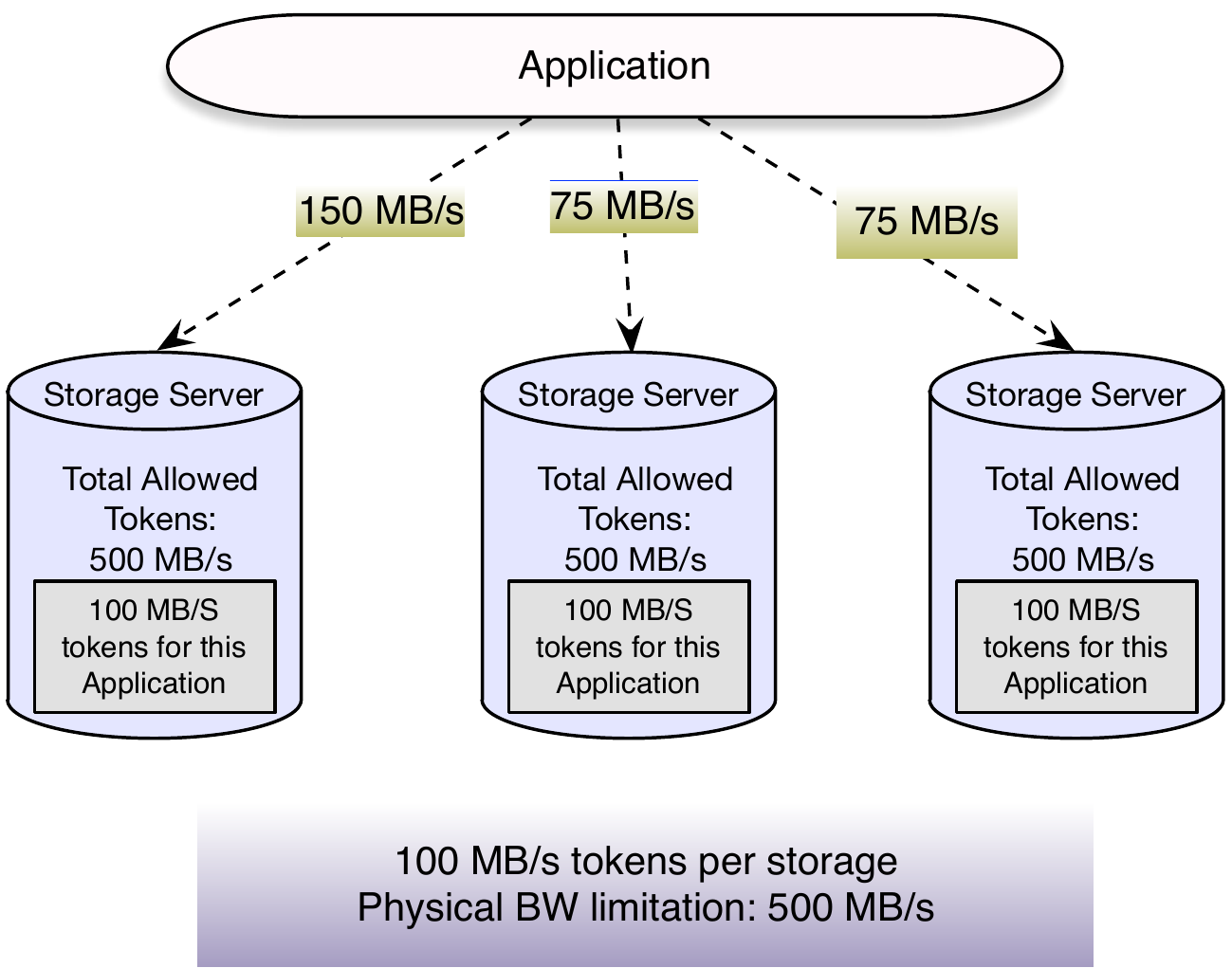}}
		\caption{Token bucket algorithm in HPC platform.}
		\label{fig22}
	\end{center}
	\vspace{-1em}
\end{figure}

In order to explain why the token bucket algorithm alone is not enough, let's consider an example. As shown in Fig.~\ref{fig22}, suppose the total physical bandwidth limit for each storage is 500 MB/s, then the application requests 300 MB/s of storage bandwidth, and the control plane assigns 300 MB/s of tokens for this application. These tokens are evenly distributed to storage servers such that each of them has 100 MB/S of tokens. 

As this application only contains 100 MB/s tokens in each storage server, server 1 could not serve requests at 150 MB/s, even though other servers actually have unused tokens for this application. In this example, only a total of 250 MB/s of bandwidth is provided even there are unused tokens for this application. To address such an issue, we propose a borrowing model described in the next section to enable the application to borrow unused tokens from other storage servers.

\subsection{Extended M-LWDF algorithm and policy enforcer}
M-LWDF~\cite{andrews2001providing} is a throughput-optimal scheduling algorithm which is used when a single shared resource is used by multiple users to satisfy each individual throughput requirements. Intuitively, the M-LWDF algorithm prioritizes the applications that are accessing the same shared resource such that they can obtain their desired bandwidth. 
Based on the M-LWDF algorithm, in each time slot $t$, application  $i$ will served if Equation~\ref{StorageMLWDF} ~\cite{andrews2001providing} is maximal (which refers to the application priority, i.e., the higher value, the higher priority for the application):
\begin{equation}
\gamma_i L_{i}(t)C_{i}(t)
\label{StorageMLWDF}
\end{equation}
in which,
\begin{equation}
\gamma_i = \frac{a_{i}}{r_i}, \\
a_i = -\frac{\log{\delta_i}}{T_i}
\end{equation}
where $L_{i}$ refers to the I/O queue length for the application $i$,  $C_{i}(t)$ is the available resource capacity  for that application at time slot $t$, and $ \gamma_i$ are arbitrary positive constants~\cite{andrews2001providing}. 
As shown in~\cite{andrews2001providing}, to satisfy a certain amount of bandwidth for applications that using a shared resource, the M-LWDF scheduling is used in conjunction with the token bucket algorithm. 

The original M-LWDF algorithm is widely used to provide a certain minimum bandwidth per application. However, it can only be used with a single shared resource (i.e., a storage server in our case). Since this study deals with multiple distributed resources, we may extend this algorithm with a borrowing model to handle distributed resources. This extension has two parts as follows:
\begin{enumerate}
	\item \emph{Distributed M-LWDF}: using the original M-LWDF from ~\cite{andrews2001providing} on each storage server.
	\item \emph{The borrowing model algorithm}: a mechanism that allows the applications to borrow unused tokens from other storage servers (i.e., a queue $Q_i$ for application $i$ to borrow tokens from the queues of the same application on other storage servers). The borrowing model is ran by the policy enforcer of the centralized control plane for each local data plane. This algorithm covers whether the borrow can happen, when, and how many tokens should be borrowed. The pesuodocode of the borrowing model on each local data plane is described in Algorithm~\ref{Borrowing Model}.
	\begin{algorithm}
		\caption{Borrowing Model Algorithm}\label{Borrowing Model}
		\begin{algorithmic}[1]
			\Procedure{Borrowing Model Algorithm}{}
			\State $\textit{Borrow = False}$;
			\State $\textit{n } \gets \text{Num of Applications }$
			\State $\textit{m } \gets \text{Num of Storage Servers }$
			\For{\texttt{$\textit{ i } \gets \text{1 to  }\textit{n}$}}
			\State $a\gets \textit{Num of assigned tokens for application}$;
			\State $d \gets \textit{Num of required tokens for application}$;
			\State $p \gets a-d$;
			\If {$p < 0$} 
			\State $\textit{Borrow = True}$;
			\State $\textit{1)randomly choose two storage servers}$;
			\State $\textit{2)select the storage server with greater}$ 
			\State $\textit{number of unsued tokens}$; 
			\State $\textit{3)Borrow}$ 
			$\bigl| p \bigr| $
			$\textit{tokens form the storage server}$
			\State $\textit{(at most)}$;
			\EndIf
			\EndFor
			\EndProcedure
		\end{algorithmic}
	\end{algorithm}	

As shown in Algorithm~\ref{Borrowing Model}, the assigned number of tokens (i.e., $a$) and the required number of tokens (i.e., $d$) are calculated and compared. For each application, located on each local data plane, if there is a need for more tokens (i.e., $a < d$), the local data plane sends a message to the centralized control plane, using the communication interface, to request more tokens. Once the centralized control plane receives a token request message, it will check other storage servers to find which ones have unused available tokens. Two of these servers will be chosen at random. Finally, among these two random choices, the storage server with the larger number of available tokens will be chosen. Note that, it is possible that the available tokens of the random selected storage is less than the required tokens. In this case the borrowing algorithm will be repeated until the centralized control plane can assign as many tokens as are required for the application. Because of this, the proposed borrowing algorithm can either satisfy the entire desired bandwidth, or at least a significant portion of it. 
\end{enumerate}	

Therefore, using the extended M-LWDF algorithm with the borrowing model introduces a dynamic number of tokens for each application. In addition, if $Q_i$ on server $S_i$ borrows extra tokens from other servers, it will gain higher priority to be served. So, the Equation~\ref{StorageMLWDF} which is used to calculate the priority is updated as follows:
\begin{equation}
\gamma_i L_{i}(t)C_{i}(t) + T
\label{NewStorageMLWDF}
\end{equation}
Where $T$ represents the total number of borrowed tokens.

In addition, we design a set of policies regarding the borrowing model as follow:
\begin{itemize}
	\item Prohibit an application to borrow tokens: \texttt{<app-i, borrow=FALSE>}.
	\item Allow an application to borrow tokens: \texttt{<app-i, borrow=TRUE>}. 
	\item Allow an application to borrow tokens if only $thres$ percent of its required bandwidth is satisfied: \texttt{<app-i, borrow=TRUE, thres=0.8>}.
\end{itemize}
The policy enforcer in the control plane will distribute the policy that users specify to control the QoS of an application.

\subsection{QoS procedure}
Suppose that  in an HPC system, several applications are concurrently running on compute nodes while each of them needs its desired I/O requirements from the storage servers. We will show how QoS can be provisioned using the proposed solution. 

\begin{enumerate}
	\item All application I/O requests and their QoS specifications are delegated to the global data plane using the communication interface.
	\item The global data plane classifies all I/O requests (based on their I/O headers) to the separate dedicated queue for each application.
	\item The global data plane uses the M-LWDF algorithm to find the priority for each application. Then, it communicates with the I/O scheduler to find appropriate storage servers for each application. Finally, the global data plane uses this information to distribute I/O requests onto the local data plane of the appropriate storage servers.
	\item The token rate generator of the centralized control plane communicates with the global data plane to get the applications' desired specifications (BW). 
	\item The token rate generator of the centralized control plane generates tokens for each application and put them into their corresponding virtual token buckets.
	\item The traffic shaper on the centralized control plane is used to evenly distribute application tokens on each storage server. 
	\item The local data plane on each storage server classifies its issued I/O requests (based on their I/O header), to the dedicated local queues per application. Then, the desired specification component of the local data plane will be updated in terms of the I/O requests for each application. For example, in the local data plane $DP_{ij}$ if the total I/O requests for application $j$ is 100 MB/s, then the desired specification (BW) for application $j$ is set to 100 MB/s. Therefore the desired number of tokens for the application $j$ on the local data plane $DP_{ij}$ is 100 tokens.
	\item On each local data plane, the extended M-LWDF algorithm is ran in such a way that:
		\begin{enumerate}
		\item  First, the borrowing algorithm is ran to compare the desired tokens with the assigned tokens for each application.
		\item  If more tokens are needed in an application, the local data plane will communicate with the centralized data plane to borrow tokens from the queues of the same application in other storage servers. 
		\item Next, a priority is calculated for each application to satisfy the entire QoS requirement of the application (or at least a proper portion of it).
		\item The borrowing algorithm is repeated until no more tokens can be borrowed.
		\end{enumerate}
\end{enumerate}

\section{Evaluation}
\label{sec:8.Evaluation}
This section introduces the performance results of our software-defined QoS provisioning framework. We evaluated our framework across two dimensions: 1) performance gain of the control plane and data plane and 2) the ability of control plane to enforce policies using the borrowing model. To reveal the actual performance of the storage servers, we generated synthetic workloads. In the following, we first explain the experimental setup which is used for this study, then we present the performance evaluation results of our framework.

\subsection{Experimental Setup}
For the experiments, we set up a number of compute nodes to send parallel read and write requests to a set of storage servers. Our testbed consists of 10 servers, each with 16  Intel Xeon 2.4 GHz E5-2665 cores and 384 GB of RAM. We assumed that the storage servers have the same configuration as described in Table~\ref{t1}. For the experiments, we first ran I/O requests of varying sizes as a test, then we ran microbenchmarks that issue I/O requests of the same size to a set of the storage servers. We repeated the experiments for I/O sizes of 4KB, 8KB, and 64KB. Our experiments address a question of how much applications bandwidth (QoS) specifications are met using the software-defined approach. Our experimental results demonstrate that using the proposed approach leads to a significant performance gain for a variety of HPC applications. The evaluation results will be discussed in the next section in detail. 

\begin{center}
	\begin{tabular}{| c | c | c | c |}
		\hline
		Storage server&  10 servers, each with 16 cores\\ \hline
		Core &  Intel Xeon 2.4 GHz E5-2665 \\ \hline
		RAM  &  384 GB of RAM  \\ \hline
	\end{tabular}
	\captionof{table}{Storage servers configuration} \label{t1} 
\end{center}

\subsection{Evaluation Results}
We evaluated the performance of our framework in terms of the allocated bandwidth (QoS) specifications for each application. Note that, bandwidth and throughput are used interchangeably in this paper. 

\begin{figure}[htbp]
	\begin{center}
		{\includegraphics[width=3.3in]{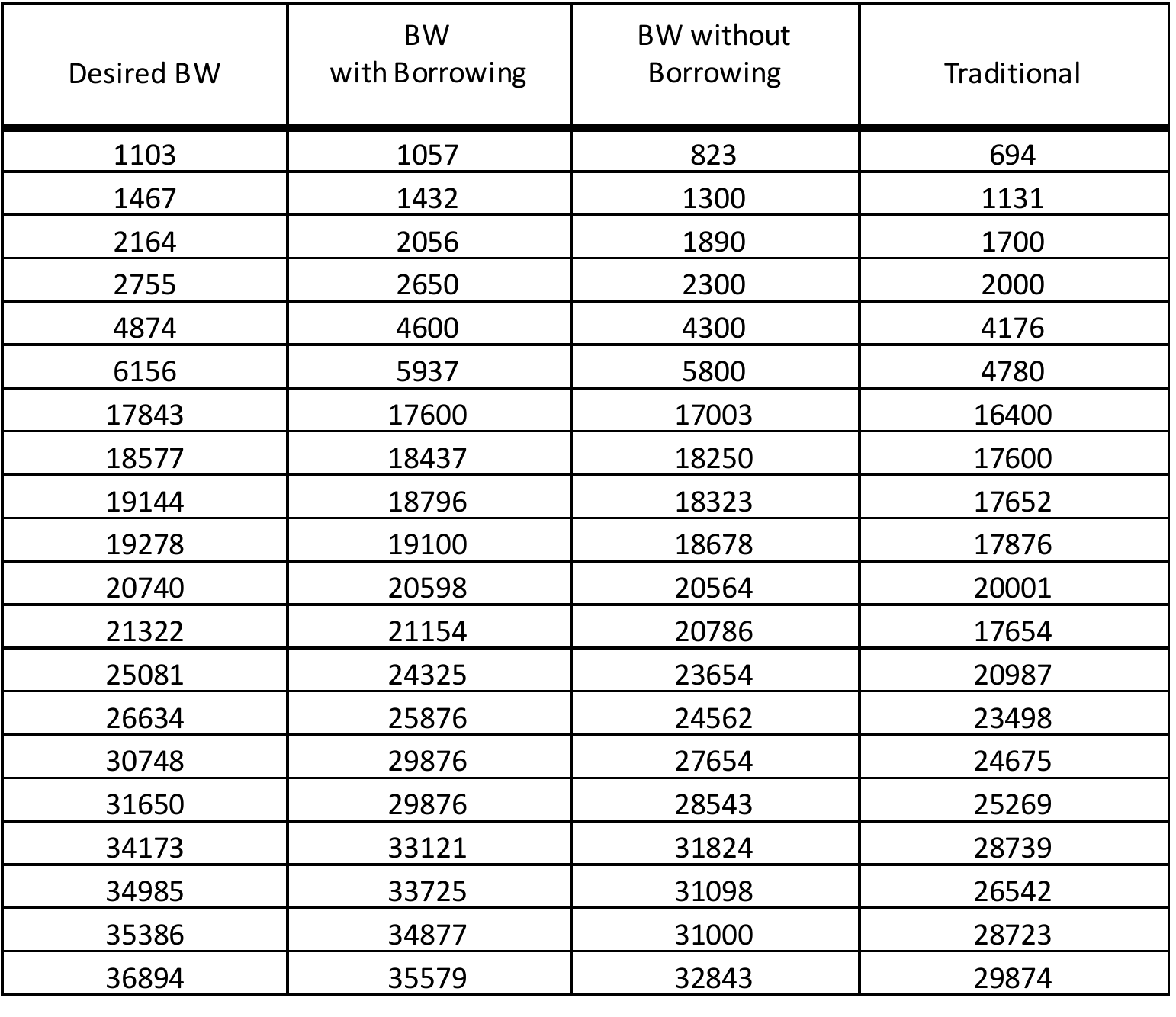}}
		\captionof{table}{Evaluation Results (Bandwidth)} \label{t2} 
	\end{center}
\end{figure}

We first ran a set of I/O requests of varying sizes as a test. Without loss of generality (and also for simplicity), we consider 20 applications which are concurrently running on a set of compute nodes, and each application has 2000 processes such that every process issues I/O requests to a set of storage servers. Workloads (i.e., I/O requests) were generated randomly using a normal distribution. 

Table~\ref{t2} shows the performance evaluation results in terms of achieved bandwidths for twenty concurrent applications which are running on ten shared storages devices. We compared our QoS provisioning framework with traditional storage framework. Note that, the traditional storage framework refers to a typical HPC framework without software-defined ability.

\begin{figure}[htbp]
	\begin{center}
		{\includegraphics[width=3.1in]{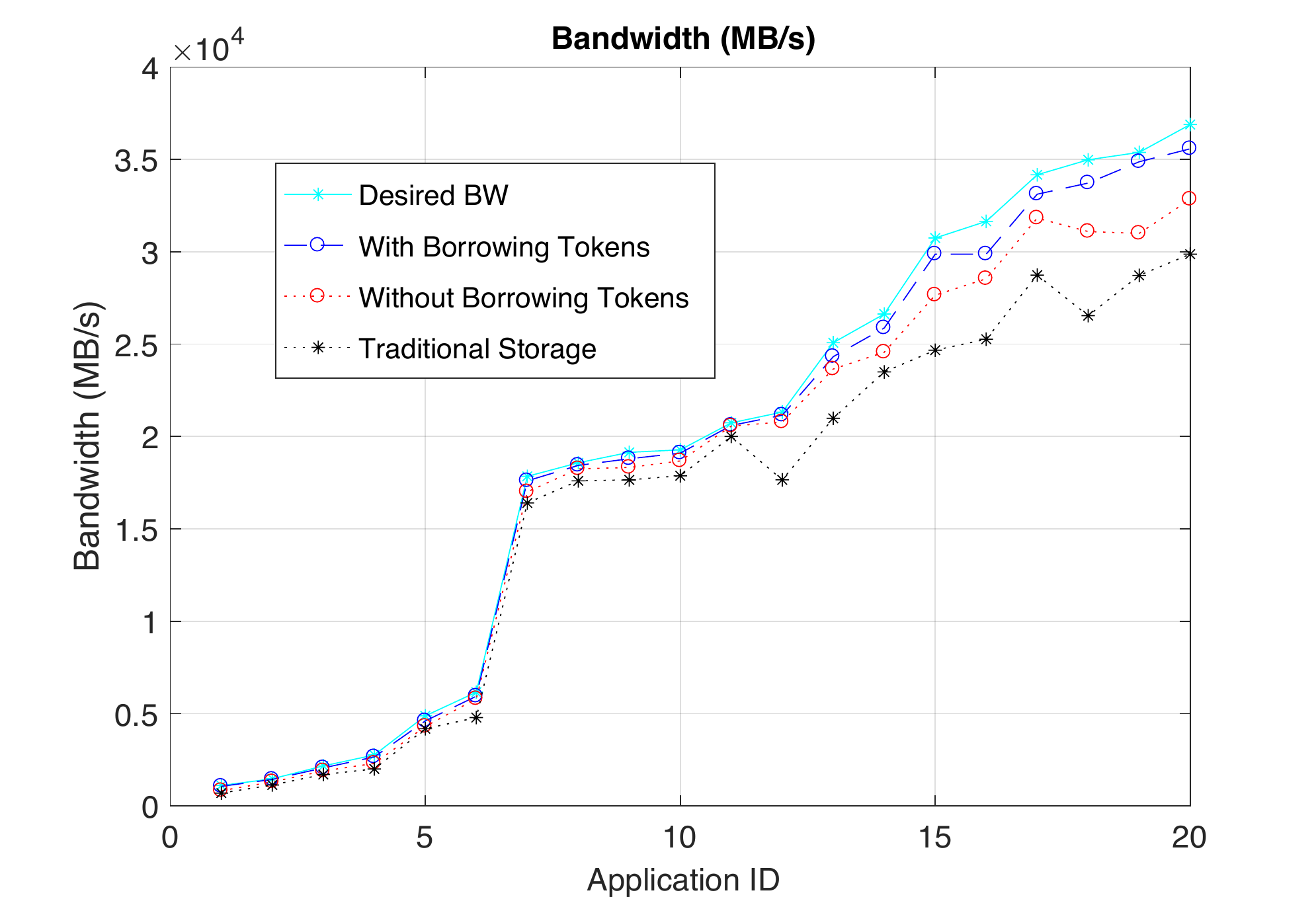}}
		\caption{Evaluation Results (Bandwidth (MB/s))}
		\label{fig3}
	\end{center}
\end{figure}

The plots of these results shown in Fig.~\ref{fig3}, which illustrates the allocated bandwidth for each application. The desired BW is compared with three scenarios: QoS-Aware with borrowing tokens, QoS-Aware without borrowing tokens, and traditional storage framework. As Fig.~\ref{fig3} shows, evaluation results demonstrate that allocated application's bandwidth for each application using our QoS provisioning framework with borrowing tokens, is close to its desired bandwidth. However, for the traditional framework, there exists a significant gap between the desired BW and the achieved bandwidth. 

Table~\ref{t3} shows allocated bandwidth percentage for applications under three scenarios, as the table shows allocated bandwidth for the borrowing model is approximately 97\% of the desired BW, however the traditional storage framework can only provide approximately 84\% of the desired BW. Therefore, our software-defined approach leads to performance gain for HPC applications. 

\begin{center}
	\begin{tabular}{| c | c | c | c |}
		\hline
		Scenario & Allocated BW (Percentage)\\ [0.5ex] 
		\hline\hline
		BW with borrowing&  97.3650422\%\\ \hline
		BW without borrowing &  92.383381\% \\ \hline
	    BW for traditional  &  84.3971722\%  \\ \hline
	\end{tabular}
	\captionof{table}{Allocated bandwidth percentage} \label{t3} 
\end{center}

The rest of this section introduces the evaluation results after running synthetic microbenchmarks that issue I/O requests of the same size to a set of storage servers. We repeated the experiments for I/O requests of size 4KB, 16KB, and 64KB and we investigated the effect of applications' I/O size on the allocated bandwidth for each application. 

Our synthetic microbenchmarks are comprised of 20 concurrent applications which are running on a set of compute nodes, each application has 2000 processes and every process issues I/O requests of the same size to a set of storage servers.  

\begin{figure}[htbp]
	\begin{center}
		{\includegraphics[width=3.1in]{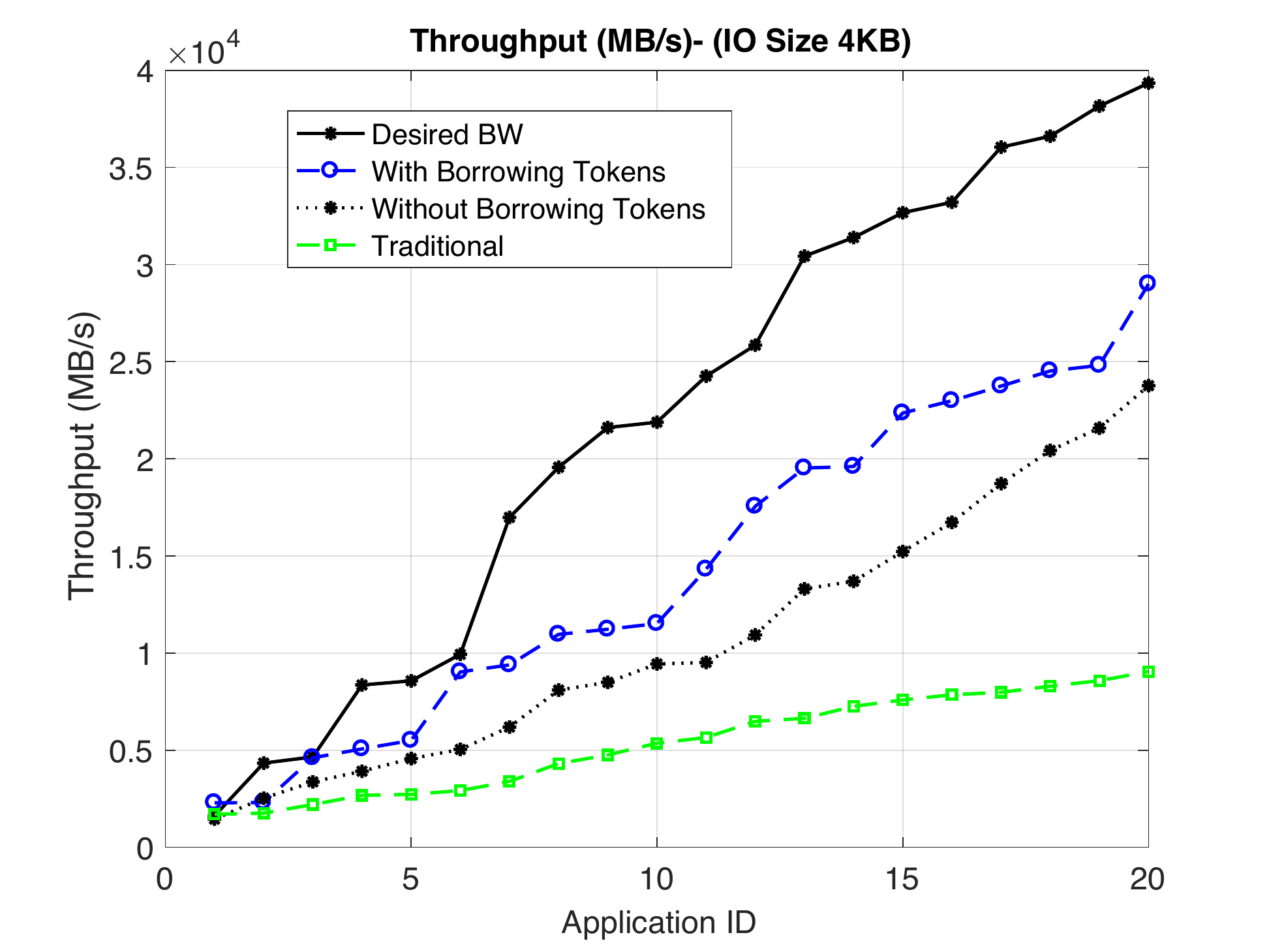} }
		\caption{Evaluation Results (Bandwidth) - Average IO size 4KB}
		\label{n1}
	\end{center}
\end{figure}

Fig.~\ref{n1} represents allocated bandwidth for each application where average IO size is 4KB. As the figure shows, using our QoS-Aware approach with borrowing token leads to decrease the gap between the allocated bandwidth for each application and its desired bandwidth. However, there is a significant gap between the desired BW and the allocated bandwidth when traditional storage framework is used.

\begin{figure}[htbp]
	\begin{center}
		{\includegraphics[width=.4\textwidth]{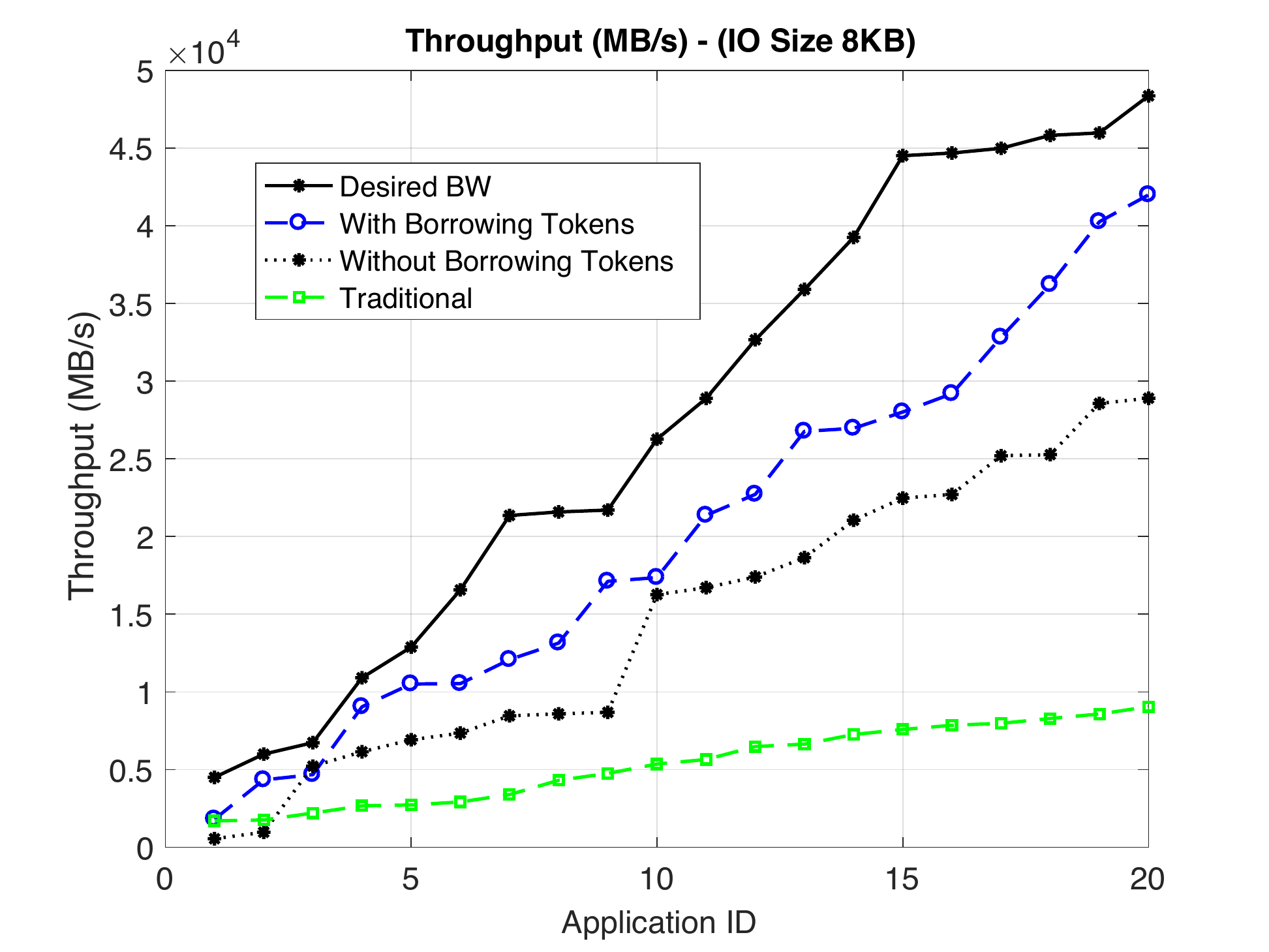} }
		\caption{Evaluation Results (Bandwidth per application- Average IO size 8KB)}
		\label{n2}
	\end{center}
\end{figure}

Table~\ref{t4} shows allocated bandwidth percentage of applications for the three aforementioned scenarios, as the table shows allocated bandwidth for the borrowing model is approximately 65\% of the desired BW, however, the traditional storage framework only can satisfy approximately 24\% of the desired BW. As a result, using our QoS-Aware approach can lead to significant performance gain for a variety of HPC applications. Such that, our platform can satisfy  65\% of total application's bandwidth which provides 41\% improvement compared to the traditional storage platform.
\begin{center}
	\begin{tabular}{| c | c | c | c |}
		\hline
		Scenario & Allocated BW (Percentage)\\ [0.5ex] 
		\hline\hline
		BW with borrowing&  65.1745806\%\\ \hline
		BW without borrowing &  48.7157613\% \\ \hline
		BW for traditional  &  24.064799\%  \\ \hline
	\end{tabular}
	\captionof{table}{Allocated bandwidth percentage} \label{t4} 
\end{center}
Fig.~\ref{n2} illustrates allocated bandwidth for each application where average IO size is 8KB. As the figure shows, our QoS-Aware framework with borrowing tokens, provides a significant bandwidth gain for each application.
Note that, based on our framework, allocated bandwidth can not be more than the desired one.

Table~\ref{t5} shows percentage of allocated bandwidth of applications for the three aforementioned scenarios, as the table shows allocated bandwidth for the borrowing model is approximately 90\% of the desired BW, however, the traditional storage framework can only provide approximately 22\% of the desired BW. Hence, using our QoS-Aware approach can lead to a significant performance gain for a variety of HPC applications. As results show, our framework can satisfy 90\% of total application's bandwidth which provides 68\% improvement compared to the traditional storage platform.

\begin{center}
	\begin{tabular}{| c | c | c | c |}
		\hline
		Scenario & Allocated BW (Percentage)\\ [0.5ex] 
		\hline\hline
		BW with borrowing&  90.1927129\%\\ \hline
		BW without borrowing &  58.8441338\% \\ \hline
		BW for traditional  &  22.0348425\%  \\ \hline
	\end{tabular}
	\captionof{table}{Allocated bandwidth percentage} \label{t5} 
\end{center}
Comparing the evaluation results of Fig~\ref{n1} and Fig~\ref{n2} demonstrates that increasing application's I/O size leads to more performance gain for HPC applications. We can hypothesise that this is due to the existence of more available tokens for performing I/O requests, as increasing the size of I/O requests leads to more available tokens for them. Hence, more tokens, leads to more performance gain at the end. However, there is a limitation for increasing application's I/O size that leads to a better performance gain.

\begin{figure}[htbp]
	\begin{center}
		{\includegraphics[width=.4\textwidth]{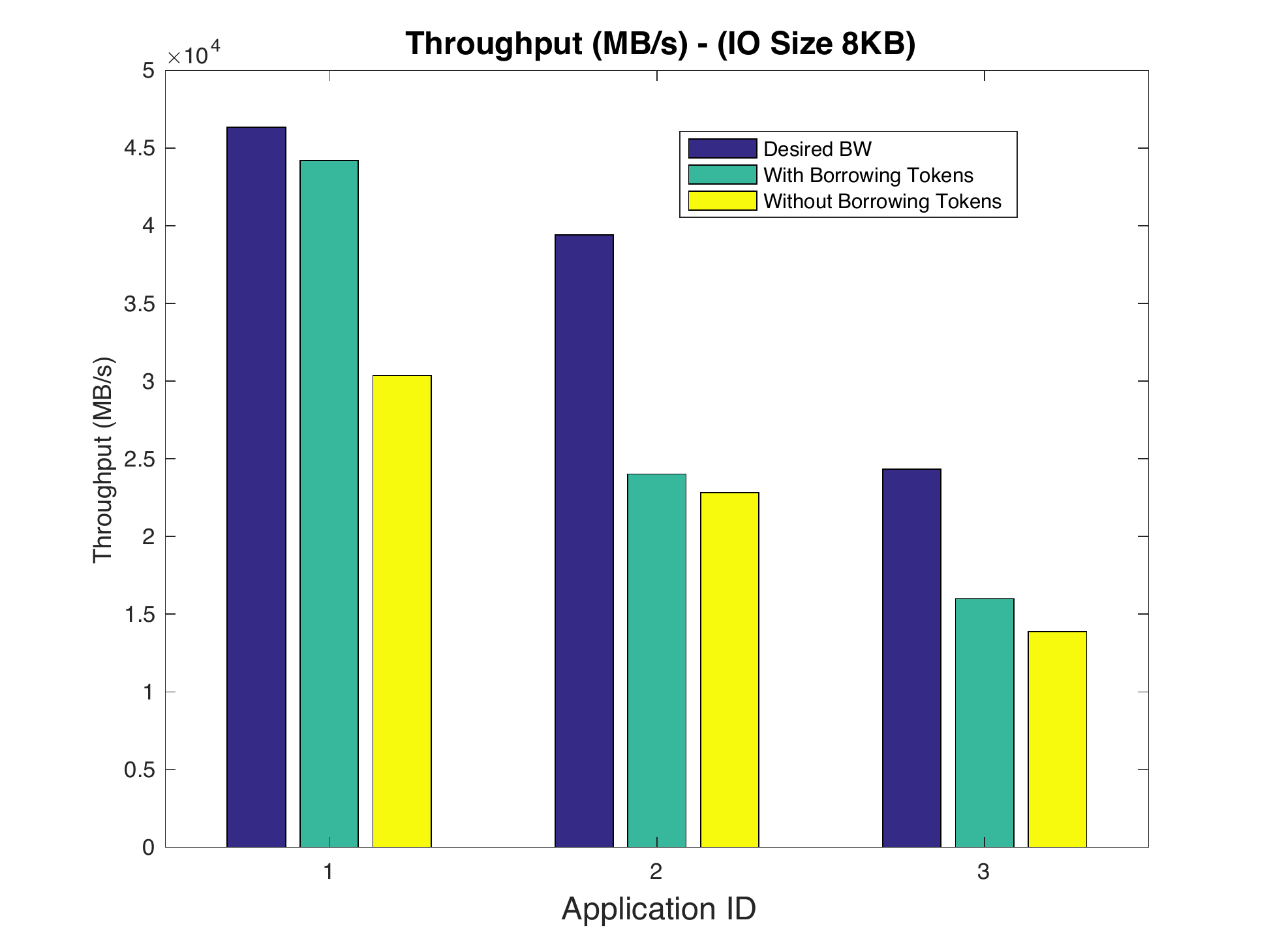} }
		\caption{Evaluation Results (Comparing total throughput of three applications}
		\label{n31}
	\end{center}
\end{figure}

Fig.~\ref{n31} represents total throughput of three random applications among 20 applications. As the figure shows, total throughput for each application using our SDS platform is close to its desired bandwidth. However, our framework can not satisfy 100\% bandwidth guaranteed for each application because of the existing physical limitations which were discussed in Section~\ref{sec:3.Background and Challenges}.

\begin{figure}[htbp]
	\begin{center}
		{\includegraphics[width=.4\textwidth]{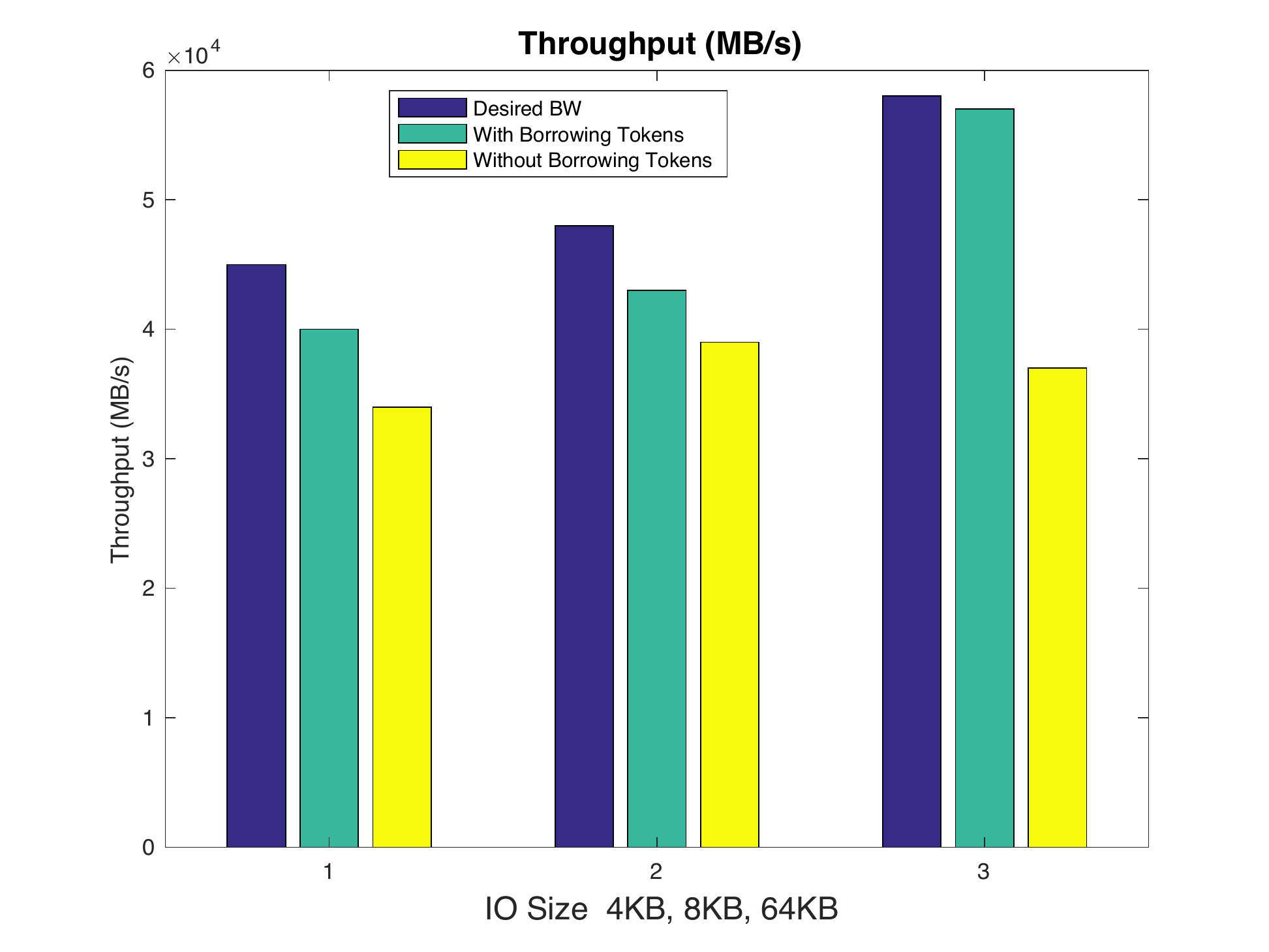} }
		\caption{Throughput vs Application I/O size}
		\label{n3}
	\end{center}
\end{figure}

Fig.~\ref{n3} illustrates effects of IO sizes on allocated throughput. As the figure shows, as the average IO size increases, the allocated throughput will be closer to that which is desired for each application. 
%One possible reason is that, ------
As the evaluation results demonstrate, deploying our QoS-Aware SDS-based approach leads to a significant performance gain for HPC applications. 

In summary, our evaluation results demonstrate that using our QoS-Aware software-defined framework can not satisfy 100\% guaranteed bandwidth for each application, however, it leads to a significant performance gain compared to the traditional HPC framework. These results were expected because of existing physical limitations that prevent to meet 100\% of the desired bandwidth for each application.

\section{Related Works}
\label{sec:4.RelatedWorks}
Existing QoS-capable storage systems can provide QoS assurance at a coarser gain, such as per application or per storage node. For instance, \emph{Zygaria}~\cite{wong2006zygaria} grantees a QoS mechanism per node. It provides QoS assurance using a hierarchical arrangement of \emph{sessions} and \emph{pools} with reserves and limits. Pool refers to a long-term entity that is created by an administrator to manage IO requests per application. In Zygaria, the amount of resources allocated for each application is specified by the administrator. \emph{Facade}~\cite{lumb2003faccade} describes an approach to providing performance guarantees in an efficient and cost-effective approach. It not only provides a fine-grain QoS enforcement, but it also adapts to changes in the application quickly. \emph{GRIO}~\cite{gildfind2011method} provides a QoS-aware framework by isolating performance between different classes of applications. It enables QoS enforcement in abject-based file system~\cite{mesnier2003object}.
In Zygaria, Facade, and GRIO, QoS assurance is satisfied per storage node, although the file systems spread are typically across multiple storage nodes. \emph{Sundaram}~\cite{sundaram2003practical} describes a system to dynamically allocate storage bandwidth to multiple applications based on a reinforcement learning approach.
Application-specific knowledge has been used to develop an efficient and practical learning-based technique for dynamic storage resource allocation. The system described in \emph{Stonehenge}~\cite{huang2003stonehenge} provides a cluster-based storage system with multi-objective QoS assurance, with metrics including bandwidth, capacity and latency. It can multiplex virtual disks with a specific bandwidth, capacity and storage for each of them while providing QoS guarantees. \emph{SLEDS}~\cite{van1998sleds} provides QoS enforcement by using special gateways between storage servers and compute nodes.
The use of per-device (or per-stream) QoS assurances is extremely common in traditional distributed storage systems, because in such environments, related data is typically stored on the same storage node.
For \emph{Ceph}~\cite{weil2006ceph}, files are broken to multiple objects distributed across different Object Server Devices (OSD), so for Ceph it is not sufficient to provide a per-device QoS. It needs a QoS mechanism for the whole system. 

Physical and mechanical features of the disks lead to a stateful nature of disk scheduling, which makes storage QoS assurance complicated. The time taken for serving an I/O request depends on the location of the I/O request as well as the current location of the disk head. Consequently, the throughput of the storage devices depends on both the workloads and the data layout. It is impossible to provide a true isolation of performance where there are shared disk arms, because different workloads that share the same disks arms might interfere with each other.
Therefore, the amount of available disk bandwidth is not fixed, which results in a different environment from the network bandwidth. Hence, the storage QoS must overcome the issue of resource allocation where the total amount of available resources varies. Storage QoS mechanisms which deal with this issue are categorized as four ways as follows:
\begin{enumerate}
	\item Proportional Sharing: In this method, each storage server gets a proportion of the disk's bandwidth. For example, in Sundaram~\cite{sundaram2003practical} the \emph{Cello}~\cite{shenoy1998cello} disk scheduler is used to allocate portions of the total disk bandwidth to different application.
	\emph{YFQ}~\cite{bruno1999disk} uses proportional sharing however, the actual bandwidth received per application varies.
	This method can not guarantee the actual defined bandwidth is satisfied. 
	\item Using Estimated Value as the total bandwidth: This estimated value is used for performance assurance. For example, \emph{DFS}~\cite{akinlar2000bandwidth}, Zygaria~\cite{wong2006zygaria} and \emph{GRIOv2}~\cite{wu2007providing} uses this estimated value as the ”qualified bandwidth": the trade-off between the QoS and the total throughput.
	\item Adaptation: This method adapts the changing of the total bandwidth. This method can be generalized into a throttling model. Throttling approach is a feedback-based model that is based on the comparison between the current bandwidth condition with that of desired condition.
	\item Profiling and Extracting of Disk Parameters: This method uses profiling and extracting of various disk parameters such as seek time, rotational latency, and access time. By using these parameters, the exact service time of a request can be calculated and will be schedule accordingly. Real time scheduler~\cite{reuther2003rotational} uses this method.	
\end{enumerate}
 
\section{Conclusion}
\label{sec:9.Conclusion}

In this paper, we present a software-defined QoS provisioning framework for HPC applications, which is a programmable framework used for QoS provisioning of HPC applications. We propose a series of novel designs into the general software-defined approach in order to deliver our goal. Specifically, we introduced a borrowing-based strategy and a new M-LWDF algorithm based on traditional token-bucket algorithms to assure a fair and efficient utilization of resources for HPC applications. Due to the lack of software-defined frameworks in current HPC platform, we evaluated our framework through simulation. The experimental results show that our strategies make a significant improvement upon the general HPC frameworks and lead to clear performance gain for HPC applications. Specifically, our evaluation results demonstrate that using our QoS-Aware software-defined framework can not satisfy 100\% guaranteed bandwidth for each application, however it leads to a significant performance gain for HPC applications compared to the traditional HPC framework. 
\section{Acknolegment}
\label{sec:10.ack}
This research is supported in part by the National Science Foundation under grant CCF-1409946 and CNS-1338078. 
%We greatly appreciate the time and efforts by the referees in reviewing this paper and the valuable suggestions offered.

\bibliographystyle{IEEEtran}
\bibliography{bib}{}

\begin{thebibliography}{10}
\providecommand{\url}[1]{#1}
\csname url@samestyle\endcsname
\providecommand{\newblock}{\relax}
\providecommand{\bibinfo}[2]{#2}
\providecommand{\BIBentrySTDinterwordspacing}{\spaceskip=0pt\relax}
\providecommand{\BIBentryALTinterwordstretchfactor}{4}
\providecommand{\BIBentryALTinterwordspacing}{\spaceskip=\fontdimen2\font plus
\BIBentryALTinterwordstretchfactor\fontdimen3\font minus
  \fontdimen4\font\relax}
\providecommand{\BIBforeignlanguage}[2]{{%
\expandafter\ifx\csname l@#1\endcsname\relax
\typeout{** WARNING: IEEEtran.bst: No hyphenation pattern has been}%
\typeout{** loaded for the language `#1'. Using the pattern for}%
\typeout{** the default language instead.}%
\else
\language=\csname l@#1\endcsname
\fi
#2}}
\providecommand{\BIBdecl}{\relax}
\BIBdecl

\bibitem{chang2008bigtable}
F.~Chang, J.~Dean, S.~Ghemawat, W.~C. Hsieh, D.~A. Wallach, M.~Burrows,
  T.~Chandra, A.~Fikes, and R.~E. Gruber, ``Bigtable: A distributed storage
  system for structured data,'' \emph{ACM Transactions on Computer Systems
  (TOCS)}, vol.~26, no.~2, p.~4, 2008.

\bibitem{krauter2002taxonomy}
K.~Krauter, R.~Buyya, and M.~Maheswaran, ``A taxonomy and survey of grid
  resource management systems for distributed computing,'' \emph{Software:
  Practice and Experience}, vol.~32, no.~2, pp. 135--164, 2002.

\bibitem{chervenak2000data}
A.~Chervenak, I.~Foster, C.~Kesselman, C.~Salisbury, and S.~Tuecke, ``The data
  grid: Towards an architecture for the distributed management and analysis of
  large scientific datasets,'' \emph{Journal of network and computer
  applications}, vol.~23, no.~3, pp. 187--200, 2000.

\bibitem{shoshani1998storage}
A.~Shoshani, L.~Bernardo, H.~Nordberg, D.~Rotem, and A.~Sim, ``Storage
  management for high energy physics applications,'' in \emph{Computing in High
  Energy Physics}, vol. 1998, 1998.

\bibitem{lantz2010network}
B.~Lantz, B.~Heller, and N.~McKeown, ``A network in a laptop: rapid prototyping
  for software-defined networks,'' in \emph{Proceedings of the 9th ACM SIGCOMM
  Workshop on Hot Topics in Networks}.\hskip 1em plus 0.5em minus 0.4em\relax
  ACM, 2010, p.~19.

\bibitem{mckeown2008openflow}
N.~McKeown, T.~Anderson, H.~Balakrishnan, G.~Parulkar, L.~Peterson, J.~Rexford,
  S.~Shenker, and J.~Turner, ``Openflow: enabling innovation in campus
  networks,'' \emph{ACM SIGCOMM Computer Communication Review}, vol.~38, no.~2,
  pp. 69--74, 2008.

\bibitem{kim2013improving}
H.~Kim and N.~Feamster, ``Improving network management with software defined
  networking,'' \emph{IEEE Communications Magazine}, vol.~51, no.~2, pp.
  114--119, 2013.

\bibitem{tang1999network}
P.~P. Tang and T.-Y. Tai, ``Network traffic characterization using token bucket
  model,'' in \emph{INFOCOM'99. Eighteenth Annual Joint Conference of the IEEE
  Computer and Communications Societies. Proceedings. IEEE}, vol.~1.\hskip 1em
  plus 0.5em minus 0.4em\relax IEEE, 1999, pp. 51--62.

\bibitem{shenker1997general}
S.~Shenker and J.~Wroclawski, ``General characterization parameters for
  integrated service network elements,'' 1997.

\bibitem{shalimov2013advanced}
A.~Shalimov, D.~Zuikov, D.~Zimarina, V.~Pashkov, and R.~Smeliansky, ``Advanced
  study of sdn/openflow controllers,'' in \emph{Proceedings of the 9th central
  \& eastern european software engineering conference in russia}.\hskip 1em
  plus 0.5em minus 0.4em\relax ACM, 2013, p.~1.

\bibitem{andrews2001providing}
M.~Andrews, K.~Kumaran, K.~Ramanan, A.~Stolyar, P.~Whiting, and R.~Vijayakumar,
  ``Providing quality of service over a shared wireless link,'' \emph{IEEE
  Communications magazine}, vol.~39, no.~2, pp. 150--154, 2001.

\bibitem{wong2006zygaria}
T.~M. Wong, R.~A. Golding, C.~Lin, and R.~A. Becker-Szendy, ``Zygaria: Storage
  performance as a managed resource,'' in \emph{Real-Time and Embedded
  Technology and Applications Symposium, 2006. Proceedings of the 12th
  IEEE}.\hskip 1em plus 0.5em minus 0.4em\relax IEEE, 2006, pp. 125--134.

\bibitem{lumb2003faccade}
C.~R. Lumb, A.~Merchant, and G.~A. Alvarez, ``Fa{\c{c}}ade: Virtual storage
  devices with performance guarantees.'' in \emph{FAST}, vol.~3, 2003, pp.
  131--144.

\bibitem{gildfind2011method}
A.~J.~A. Gildfind and K.~J. McDonell, ``Method for empirically determining a
  qualified bandwidth of file storage for a shared filed system using a
  guaranteed rate i/o (grio) or non-grio process,'' Mar.~15 2011, uS Patent
  7,908,410.

\bibitem{mesnier2003object}
M.~Mesnier, G.~R. Ganger, and E.~Riedel, ``Object-based storage,'' \emph{IEEE
  Communications Magazine}, vol.~41, no.~8, pp. 84--90, 2003.

\bibitem{sundaram2003practical}
V.~Sundaram and P.~Shenoy, ``A practical learning-based approach for dynamic
  storage bandwidth allocation,'' in \emph{Quality of Service—IWQoS
  2003}.\hskip 1em plus 0.5em minus 0.4em\relax Springer, 2003, pp. 479--497.

\bibitem{huang2003stonehenge}
L.~Huang, ``Stonehenge: A high performance virtualized storage cluster with qos
  guarantees,'' 2003.

\bibitem{van1998sleds}
R.~Van~Meter, ``Sleds: storage latency estimation descriptors,'' in \emph{IEEE
  Symposium on Mass Storage Systems}, 1998.

\bibitem{weil2006ceph}
S.~A. Weil, S.~A. Brandt, E.~L. Miller, D.~D. Long, and C.~Maltzahn, ``Ceph: A
  scalable, high-performance distributed file system,'' in \emph{Proceedings of
  the 7th symposium on Operating systems design and implementation}.\hskip 1em
  plus 0.5em minus 0.4em\relax USENIX Association, 2006, pp. 307--320.

\bibitem{shenoy1998cello}
P.~J. Shenoy and H.~M. Vin, ``Cello: a disk scheduling framework for next
  generation operating systems,'' in \emph{ACM SIGMETRICS Performance
  Evaluation Review}, vol.~26, no.~1.\hskip 1em plus 0.5em minus 0.4em\relax
  ACM, 1998, pp. 44--55.

\bibitem{bruno1999disk}
J.~Bruno, J.~Brustoloni, E.~Gabber, B.~Ozden, and A.~Silberschatz, ``Disk
  scheduling with quality of service guarantees,'' in \emph{Multimedia
  Computing and Systems, 1999. IEEE International Conference on}, vol.~2.\hskip
  1em plus 0.5em minus 0.4em\relax IEEE, 1999, pp. 400--405.

\bibitem{akinlar2000bandwidth}
C.~Akinlar and S.~Mukherjee, ``Bandwidth guarantee in a distributed multimedia
  file system using network attached autonomous disks,'' in \emph{Real-Time
  Technology and Applications Symposium, 2000. RTAS 2000. Proceedings. Sixth
  IEEE}.\hskip 1em plus 0.5em minus 0.4em\relax IEEE, 2000, pp. 237--246.

\bibitem{wu2007providing}
J.~C. Wu and S.~A. Brandt, ``Providing quality of service support in
  object-based file system.'' in \emph{MSST}, vol.~7, 2007, pp. 157--170.

\bibitem{reuther2003rotational}
L.~Reuther and M.~Pohlack, ``Rotational-position-aware real-time disk
  scheduling using a dynamic active subset (das),'' in \emph{Real-Time Systems
  Symposium, 2003. RTSS 2003. 24th IEEE}.\hskip 1em plus 0.5em minus
  0.4em\relax IEEE, 2003, pp. 374--385.

\end{thebibliography}

\end{document}